\documentclass[12pt,aps,prd,preprint,tightenlines,superscriptaddress,
   showpacs,nofootinbib]{revtex4}
\newcommand{\PRE}[1]{{#1}} 
\usepackage{bm} \usepackage{epsfig}

\newcommand{\sigmaSI}{\sigma_{{\rm SI}}}

\newcommand{\gev}{{\rm GeV}}

\newcommand{\pb}{{\rm pb}}
\newcommand{\cm}{{\rm cm}}
\newcommand{\m}{{\rm m}}

\newcommand{\s}{{\rm s}}

\newcommand{\Gyr}{{\rm Gyr}}

\newcommand{\secref}[1]{Sec.~\ref{sec:#1}}

\newcommand{\figref}[1]{Fig.~\ref{fig:#1}}

\newcommand{\eg}{{\em e.g.}}
\newcommand{\gsim}{\lower.7ex\hbox{$\;\stackrel{\textstyle>}{\sim}\;$}}

\begin{document}

\preprint{UH-511-1137-09}

\title{
\PRE{\vspace*{1.5in}}
From DAMA/LIBRA To Super-Kamiokande
\PRE{\vspace*{0.3in}}
}

\author{Jason Kumar%
}
\address{Department of Physics and Astronomy, University of
Hawai'i, Honolulu, HI 96822, USA
}

\begin{abstract}
We consider the prospects for probing low-mass dark
matter with the
Super-Kamiokande experiment.
We show
that upcoming analyses including fully-contained events with
sensitivity to dark matter masses from 5 to 10 GeV can test
the dark matter interpretation of the
DAMA/LIBRA signal.  We consider prospects of this
analysis for two light dark matter candidates:
neutralinos and  WIMPless dark matter.
\end{abstract}




\maketitle

\section{Introduction}

The DAMA/LIBRA experiment has seen, with $8.2\sigma$
significance~\cite{Bernabei:2008yi}, an annual
modulation~\cite{Drukier:1986tm} in the rate of scattering events,
which could be consistent with dark matter-nucleon scattering.  Much
of the region of dark matter parameter space that is favored by DAMA
is excluded by null results from other direct detection experiments,
including CRESST~\cite{Angloher:2002in}, CDMS~\cite{Akerib:2005kh},
XENON10~\cite{Angle:2007uj}, TEXONO~\cite{Lin:2007ka,Avignone:2008xc},
and CoGeNT~\cite{Aalseth:2008rx}.  On the other hand, astrophysical
uncertainties~\cite{Brhlik:1999tt,Gondolo:2005hh} and detector
effects~\cite{Bernabei:2007hw} may open up regions that can
simultaneously accommodate the results from DAMA and these other
experiments.

\subsection{A New Window for Light Dark Matter}

The sensitivity of direct detection experiments is largely determined
by the kinematics of non-relativistic elastic nuclear scattering.
If the nucleus recoil energy is measured, the momentum
transfer can then determined, which in turn determines the dark
matter mass (for a standard dark matter velocity distribution).
This connection allows a direct detection
experiment to convert an observed event rate into a detection
of dark matter with a particular mass, or alternatively, convert a
lack of a signal into an exclusion of a region of
$(m_{DM}, \sigmaSI)$ parameter space.

But the procedure described above implicitly contains the uncertainties
which could potentially lead to consistency between the dark matter
interpretation of the DAMA/LIBRA signal and the lack of signal at
other detection experiments.  All direct detection experiments have
a recoil energy threshold; nuclei recoiling with energies below this
threshold cannot be detected.  Low nuclei recoil energies imply
a low dark matter mass.  Thus,
the sensitivity of direct detection experiments tends to suffer at low
mass, as seen if Fig. \ref{fig:supdirect}.  DAMA has a
recoil energy threshold which is lower than many other direct detection
experiments, thus raising the possibility that its signal is a
result of elastic scatter with a light dark matter candidate.

However, there are some direct detection experiments,
such as TEXONO~\cite{Lin:2007ka,Avignone:2008xc}
and CoGeNT~\cite{Aalseth:2008rx},
which also have very low energy thresholds, and whose
negative results would seem to exclude the light dark matter region
which would be preferred by the DAMA signal.  But there are more
uncertainties at issue.  One example is the measurement of the recoil
energy.  DAMA uses a NaI crystalline scintillator which converts
the recoil energy of the nuclei into light, which is measured.
But, a fraction of the recoil energy is transferred to phonons
in the lattice and lost.  So the measured energy
must be scaled by a quenching factor to determine the actual
nucleus recoil energy.  But the DAMA experiment has recently
noted the existence of the ``channeling effect," a property of
crystalline scintillators wherein a ion moving through certain channels
in the lattice will lose no energy to phonons.
In this case, scaling by the old quenching
factor would overestimate the recoil energy,  and thus
the dark matter mass.  The DAMA collaboration has
adjusted its analysis to account for this effect, and
their resulting preferred parameter region is not excluded by
any other experiment~\cite{Bernabei:2007hw,Petriello:2008jj}.

On the other hand, any uncertainty in the dark matter
velocity distribution
would alter the
dark matter mass which should be inferred from any particular
measured recoil energy distribution.  If local streams of dark matter
alter the velocity distribution from that expected in our local
neighborhood, then this could also potentially shift the mass region
preferred by the
DAMA recoil energy signal~\cite{Brhlik:1999tt,Gondolo:2005hh}.

These effects are detailed in Fig. \ref{fig:supdirect}.
The application of both these effects to the DAMA
signal is somewhat controversial.  It is not at all clear that a
dark matter stream with sufficient velocity to shift the DAMA preferred
region appreciably is reasonable from the point of view of astrophysics.
And the effect of channeling on a scintillator like DAMA for the
$1-10~\gev$ energy range is currently being studied and cross-checked
by various groups.

As if this controversy were not enough, three
groups~\cite{Chang:2008xa,Fairbairn:2008gz,Savage:2008er}
have analyzed the spectrum
of modulations within the recoil energy bins which DAMA reports.
Naturally, they all reach different conclusions, ranging from ruling
out the dark matter interpretation of DAMA to declaring it completely
consistent with the spectrum of the DAMA signal.

It is far from clear what DAMA is actually seeing.
What is clear, however, is that if DAMA is
seeing dark matter, one preferred region of parameter space has dark
matter mass in the range $m_X \sim 1-10~\gev$ and spin-independent
proton scattering cross section $\sigmaSI \sim 10^{-5} - 10^{-2}~\pb$.
This is a mass region where several different experimental effects
can push in different directions,
and potentially create a window where dark
matter could be observed at DAMA while not being ruled out by
other experiments.

Moreover, there are a variety of theoretical models which attempt
to explore this region of parameter space.
Although neutralinos have been proposed as an
explanation~\cite{Bottino:2003iu}, such low masses and high cross
sections are not typical of weakly-interacting massive particles
(WIMPs), and alternative candidates have been suggested to
explain the DAMA signal~\cite{Smith:2001hy,Feng:2008ya,Feng:2008mu,%
Foot:2008nw,Feng:2008dz,Khlopov:2008ty,Andreas:2008xy,Dudas:2008eq,Kim:2009ke}.

\subsection{Cross-checking DAMA}

The current state of affairs also makes it abundantly clear that
complementary experiments are likely required to sort out the true
nature of this result.  Other direct detection
experiments may play this role.  In this work, we note that
corroborating evidence may come from a very different source, namely,
from the indirect detection of dark matter at Super-Kamiokande
(Super-K)~\cite{Feng:2008qn}.  In contrast to 
direct detection experiments, which rapidly
lose sensitivity at low masses,
Super-K's limits remain strong for low masses.
But in contrast to other indirect detection experiments, which can
only be compared to DAMA after making astrophysical assumptions which
are highly uncertain (such as the cuspiness of the dark matter density
profile near the galactic center), Super-K offers a way of testing
the DAMA result which is largely model independent.
Super-K is therefore poised as one of the most promising experiments
to either corroborate or exclude many dark matter interpretations of
the DAMA/LIBRA data.

In \secref{relating}, we show the relation between
the DAMA and Super-K event rates.
In \secref{projection}, we show that
there is significant potential for Super-K to extend its reach to dark
matter masses from 5 to 20 GeV and provide sensitivity that is
competitive with, or possibly much better than, direct detection
experiments.  In \secref{models}, we apply our analysis to two
specific dark matter candidates that have been proposed to explain
DAMA: neutralinos~\cite{Bottino:2003iu} and WIMPless dark
matter~\cite{Feng:2008ya,Feng:2008dz,Feng:2008mu}.
We present our conclusions in
\secref{summary}.

\section{Bounding $\sigma_{SI}$ With Super-Kamiokande}
\label{sec:relating}

Super-K can probe dark
matter in the Sun or Earth's core annihilating
to standard model (SM)
particles, which subsequently emit neutrinos.  Muon
neutrinos then interact weakly at or near the detector to
produce muons, which are detected at Super-K.  The observed rate of
upward-going muon events places an upper bound on the
annihilation rate of dark matter in the Sun or the Earth's core.
For low-mass dark matter, the dominant contribution to neutrino
production via dark matter annihilation is from the
Sun~\cite{Gould:1987ir}, on which we focus.

The total annihilation rate is
\begin{equation}
\Gamma = {1\over 2} C \tanh^2 [ (aC)^{\frac{1}{2}} \tau ] \ ,
\end{equation}
where $C$ is the capture rate, $\tau \simeq 4.5~\Gyr$ is the age of
the solar system, and $a= \langle \sigma_{ann.} v
\rangle /(4\sqrt{2} V)$,
with $\sigma_{ann.}$ the total dark matter annihilation cross section and $V$
the effective volume of WIMPs in the Sun
($V = 5.7 \times 10^{30}\,{\rm cm}^3 (1\,{\rm GeV} /
m_X)^{3/2}$)~\cite{Gould:1987ir,Gould:1987ju,Hooper:2008cf}.
If $\langle \sigma_{ann.} v \rangle \sim 10^{-26}~\cm^3~\s^{-1}$ (to get
the observed dark matter relic density), then for the range
of parameters considered here, the Sun is in
equilibrium~\cite{Griest:1986yu,Gould:1987ir,Kamionkowski:1994dp}
and $\Gamma \approx {1\over 2}C$.
WIMP evaporation is not relevant
if $m_X \gsim 4\,{\rm GeV}$~\cite{Griest:1986yu,Gould:1987ju,Hooper:2008cf}.

The dark matter capture rate is~\cite{Gould:1987ir}
\begin{equation}
C = \left[ \left( {8\over 3\pi} \right)^{1\over 2} \sigma
{\rho_X \over m_X} \bar v   {M_B \over m} \right]
\left[ {3\over 2} {\langle v^2 \rangle \over
\bar v^2} \right] f_2 f_3 \ .
\label{capturerate}
\end{equation}
The first bracketed factor counts the rate of dark matter-nucleus
interactions: $\sigma$ is the dark matter-nucleus scattering cross
section, $\rho_X/m_X$ is the local dark matter number density, $m$ is
the mass of the nucleus, and $M_B$ is the mass of the capturing
object.  The velocity dispersion of the dark matter is $\bar v$, and
$\langle v^2 \rangle$ is the squared escape velocity averaged
throughout the Sun.  The second bracketed expression is the
``focusing'' factor that accounts for the likelihood that a scattering
event will cause the dark matter particle to be captured.  The
parameters $f_2$ and $f_3$ are computable ${\cal O}(1)$ suppression
factors that account for the motion of the Sun and the mismatch
between $X$ and nucleus masses, respectively.
$f_3 \sim 1$ for solar capture~\cite{Gould:1987ir}.
The capture rate is thus a completely computable
function of $\sigma/m_X$.  Assuming $\rho_X =
0.3~\gev~\cm^{-3}$, $\bar v \sim 300{{\rm km}
\over {\rm s}}$, ${3\over 2} {\langle v^2 \rangle \over
\bar v^2} \sim 20 $~\cite{Gould:1987ir}, and
taking  $f_2 \sim f_3 \sim 1$,
one finds $C \sim
10^{29}~(\sigma/m_X)~\gev~\pb^{-1}~\s^{-1}$.

The major remaining particle physics uncertainty is the neutrino
spectrum that arises from dark matter annihilation.
Assuming
the dark matter annihilates only to SM particles, a conservative
estimate for neutrino production may be obtained by assuming that the
annihilation to SM particles is dominated by $b \bar b$ production for
$m_b < m_X < M_W$, by $\tau \bar \tau$ production for $m_W < m_X <
m_t$, and by $W,Z$ production for $m_X > m_t$~\cite{Jungman:1994jr}.

\begin{figure}
\psfig{file=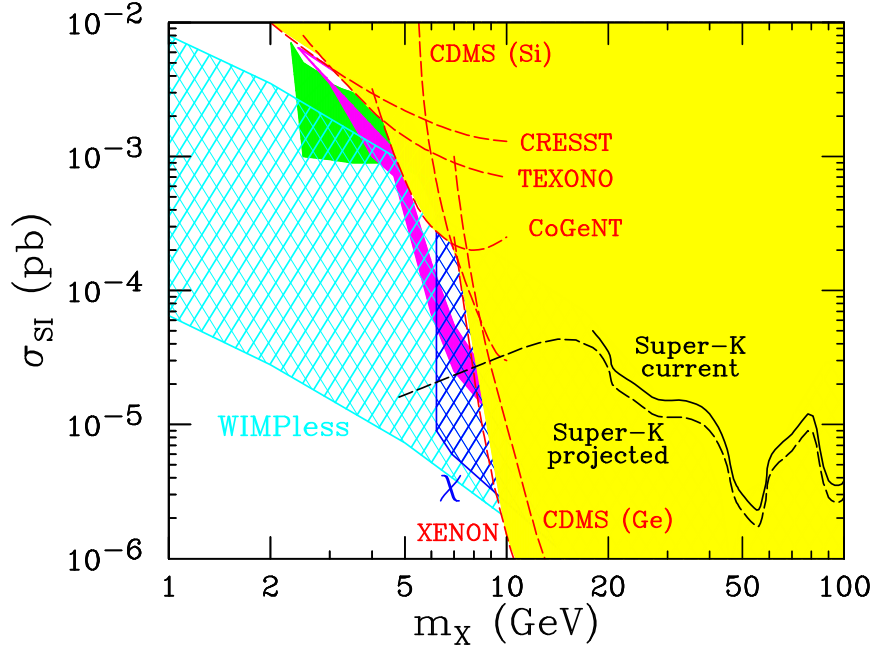,width=4.5in}
\caption{Direct detection cross sections for spin-independent
$X$-nucleon scattering as a function of dark matter mass $m_X$.  The
black solid line is the published Super-K exclusion
limit~\cite{Desai:2004pq}, and the black dashed line is our projection
of future Super-K sensitivity. The magenta shaded region is
DAMA-favored given channeling and no streams~\cite{Petriello:2008jj}, and
the medium green shaded region is DAMA-favored at 3$\sigma$ given
streams but no channeling~\cite{Gondolo:2005hh}.  The light yellow
shaded region is excluded by the direct detection experiments
indicated.  The dark blue cross-hatched
region is the prediction for
the neutralino models~\cite{Bottino:2003iu} and the
light blue cross-hatched region is the parameter space of WIMPless
models with connector quark mass $m_Y = 400~\gev$ and $0.3 <\lambda_b
< 1.0$.
Other limits come from the Baksan and MACRO experiments
\cite{Montaruli:1999nv,delosHeros:2007hy,Desai:2004pq}, though
they are not as sensitive as Super-K.}
\label{fig:supdirect}
\end{figure}

Super-K bounds the $\nu_{\mu}$-flux from dark matter annihilation in
the Sun.  Since the total annihilation rate is equal to the capture
rate, this permits Super-K to bound the dark matter-nucleon scattering
cross section using Eq. (\ref{capturerate}).
\figref{supdirect} shows the published bounds
from Super-K, limits from other dark matter direct
detection experiments and the regions of $(m_X, \sigmaSI)$ parameter
space favored by the DAMA signal given astrophysical and
detector uncertainties.  As evident from \figref{supdirect}, the
published Super-K bounds (solid line) do not test the DAMA-favored
regions.  But we will see that
consideration of the full Super-K event sample provides significant
improvement and extends Super-K's sensitivity to low masses and the
DAMA-favored regions.

\section{Projection of Super-K Sensitivity}
\label{sec:projection}

As shown in \figref{supdirect}, Super-K currently reports dark matter
bounds only down to $m_X = 18~\gev$.
For heavier dark matter, it is estimated~\cite{Desai:2004pq}
that more than 90\% of the upward-going muons will be
through-going (i.e., will be created outside the inner
detector and will pass
all the way through it).  However, one can study dark matter at lower masses
by using stopping, partially contained, or fully contained muons, that
is, upward-going muons that stop within the detector, begin within the
detector, or both\footnote{Testing the dark matter interpretation of
DAMA has with Super-K through-going muon data has also been
considered~\cite{Hooper:2008cf}.}.

Our strategy for projecting dark matter bounds from these event
topologies is as follows:
we begin by
conservatively assuming that the measured neutrino spectrum at low
energies matches the predicted atmospheric background.  In any given
bin with $N$ neutrino events, the $2\sigma$ bound on the number of neutrinos
from dark matter annihilation is then $2\sqrt{N}$.  This
bounds the dark matter annihilation rate to neutrinos, which, for a
conservative choice of the neutrino spectrum, implies
a bound on the capture rate,
and thus on $\sigma / m_X$.  To include experimental
acceptances and efficiencies, we scale our results to the Super-K
published
bound at $m_X = 18~\gev$, assuming these effects do not vary greatly
in extrapolating down to the $5-10~\gev$ range of interest.

The annihilation of dark matter particles $X$ typically
produces neutrinos with
$E_{\nu} / m_X \sim {1\over 3} - {1\over 2} $.
The muons produced by weak interactions of $\nu_{\mu}$
lie in a cone around the direction to the Sun with rms
half-angle of approximately
$\theta = 20^\circ
\sqrt{10~\gev/E_{\nu}}$~\cite{Jungman:1995df}.
Bounds on dark matter with $m_X = 18~\gev$
were set using neutrinos with
energies $E_{\nu} \sim 6-9~\gev$~\cite{Desai:2004pq}.  The
event sample used consisted of 81 upward through-going muons within a
$22^{\circ}$ angle of the Sun collected from 1679 live days.

For masses $m_X \sim 5-10~\gev$, the $\nu_{\mu}$ are
typically produced with
energies between $2-4~\gev$.  At these energies the
detected events are dominantly fully-contained events~\cite{Ashie:2005ik},
so we use this event topology with only events
within the required cone around the Sun.  The
number of events is
\begin{equation}
N_{solar} = N \frac{1-\cos\theta}{2} \ ,
\end{equation}
where $N$ is the total number of fully-contained muon events
expected in the
$2-4\,{\rm GeV}$ energy range and $\theta$ is the cone
opening angle.  Super-K expects $N_{solar} = 168$ such
fully contained events per 1000 live days~\cite{Ashie:2005ik}.

We convert this limit on event rate to a limit on the neutrino
flux by dividing by the effective cross section for the Super-K
experiment in the relevant energy range.
The effective cross-section
can be estimated by dividing the
estimated rate of events by the predicted atmospheric flux, integrated
over the relevant
range of energies~\cite{Ashie:2005ik}.  For fully contained events with
$E_{\nu} \sim 2-4~\gev$, the effective
cross section
is $\sim 2.1 \times 10^{-8}~\m^2$.  For upward
through-going events with $E_{\nu}\sim 8~\gev$, the
effective cross-section is $\sim 1.7 \times 10^{-8}~\m^2$.

Assuming the neutrino events are detected primarily in either the
fully-contained ($2-4~\gev$) or through-going
sample ($\sim 8~\gev$), one can set
$2\sigma$ limits on the time-integrated neutrino flux
due to dark matter annihilation:
\begin{eqnarray}
\Phi_{{\rm FC}}^{{\rm max}} &=& {2 \sqrt{N_{{\rm FC}}}
\over 2.1 \times 10^{-8}~\m^2 } \sim 1.6 \times 10^{9}~\m^{-2} \,
\sqrt{\frac{N_{{\rm days}}}{1679}}
\nonumber\\
\Phi_{{\rm TG}}^{{\rm max}} &=& {2 \sqrt{N_{{\rm TG}}}
\over 1.7 \times 10^{-8}~\m^2 } \sim 1.0 \times 10^{9}~\m^{-2} \,
\sqrt{\frac{N_{{\rm days}}}{1679}} \ ,
\end{eqnarray}
where $N_{{\rm FC}} = 168\, (N_{{\rm days}} / 1000)$ and
$N_{{\rm TG}} = 81\, (N_{{\rm days}} / 1679)$ are the number of
fully-contained and through-going events within the angle and energy
ranges, respectively, scaled to $N_{{\rm days}}$ live days.

The ratio of these flux limits obtained from the fully-contained and
through-going samples are then equal to the ratio of $\sigma / m_X$ in
the $5-10~\gev$ regime to the same quantity at $18~\gev$.  We find
\begin{equation}
{1.6 \times 10^{9}~\m^{-2} \over 1.0 \times 10^{9}~\m^{-2}}
\sim
\left({\sigma_{5-10} \over m_{5-10}}\right) \left({\sigma_{18}
\over 18~\gev}\right)^{-1} \ ,
\end{equation}
where $\sigma_{5-10}$ is the Super-K bound on the dark matter nucleon
cross-section for a dark matter particle with mass in the range
$5-10~\gev$, and $\sigma_{18}$ is the bound for a dark matter particle
with mass $18~\gev$.  In \figref{supdirect} this projected Super-K
bound is plotted, assuming 3000 live days of the SK I-III run.
This bound gets better at lower energies and may beat other
direct detection experiments.

\section{Prospects for Various Dark Matter Candidates}
\label{sec:models}

We now consider specific examples of theoretical models that have been
proposed to explain the DAMA result.
We first consider neutralino dark matter.
Although neutralinos typically have larger masses and lower cross
sections than required to explain the DAMA signal, special choices of
supersymmetry parameters may yield values in the DAMA-favored
region~\cite{Bottino:2003iu}.

The region of the $(m_X, \sigmaSI)$ plane spanned by these
models~\cite{Bottino:2003iu} that do not violate known constraints is
given in \figref{supdirect}.    We see that if Super-K's limits can be
extended to lower mass, it could find evidence for models in this class.
Note, however, that many of these models have $\rho
< 0.3~\gev~\cm^{-3}$, and for these models Super-K's bound on the cross
section will be less sensitive.

WIMPless dark matter provides an alternative explanation of the
DAMA/LIBRA signal~\cite{Feng:2008dz}.  These candidates are hidden
sector particles that naturally have the correct relic
density~\cite{Feng:2008ya}.  In these models, the dark matter particle
$X$ couples to SM quarks via exchange of a particle $Y$ that
is similar to a 4th generation quark.  The Lagrangian for this
interaction is
\begin{equation}
{\cal L} = \lambda_f X \bar{Y}_L f_L
+ \lambda_f X \bar{Y}_R f_R \ .
\label{connector}
\end{equation}
The Yukawa couplings $\lambda_f$ are model-dependent, and it is
assumed that only the coupling to 3rd generation quarks is
significant, while the others are Cabbibo-suppressed.\footnote{This is
a reasonable assumption and is consistent with small observed
flavor-changing neutral currents.}  One finds that the
dominant nuclear coupling of WIMPless dark matter is to gluons via a
loop of $b$-quarks ($t$-quark loops are suppressed by $m_t$).  The
$X$-nucleus cross section is given by~\cite{Feng:2008dz}
\begin{equation}
\sigmaSI = \frac{1}{4\pi}
\frac{m_N^2}{(m_N + m_X)^2}
\left[ \sum_q \frac{\lambda_b^2}{m_Y - m_X}
\left[ Z B^p_b + (A-Z) B^n_b \right] \right]^2 \ ,
\end{equation}
where $Z$ and $A$ are the atomic number and mass of the target nucleus
$N$, and $B^{p,n}_b = (2/27) m_{p} f^{p,n}_g / m_b$, where $f^{p,n}_g
\simeq 0.8$~\cite{Cheng:1988im,Ellis:2001hv}.

In \figref{supdirect}, we plot the parameter
space for WIMPless models with $m_Y = 400~\gev$ and $0.3 <\lambda_b <
1.0$.  These models span a large range in
the $(m_X, \sigmaSI)$ plane, and overlap much of the
DAMA-favored region.
We see that Super-K's projected sensitivity may
be sufficient to discover a signal that corroborates DAMA's.
But WIMPless models illustrate an important caveat to the analysis
above; if there are hidden decay channels, then the
annihilation rate to SM particles is only a fraction
of $\Gamma_{tot}$, and Super-K's sensitivity is reduced
accordingly.

\section{Summary}
\label{sec:summary}

The DAMA/LIBRA signal has focussed attention on the possibility of
light dark matter, and alternative
methods for corroborating or excluding a dark matter interpretation
are desired.  We have shown that Super-K, through its
search for dark matter
annihilation to neutrinos, has promising prospects for testing
DAMA at low mass.

Using fully contained muon events, we expect that current
super-K bounds may be extended down to $M_{DM}\sim 5-10~\gev$,
and can test light dark models (such as
neutralino models~\cite{Bottino:2003iu} and
WIMPless models~\cite{Feng:2008dz}).
We have the intriguing prospect that the
DAMA/LIBRA signal could be sharply tested by an indirect detection
experiment in the near future.

\section*{Acknowledgments}

We are grateful to Hank Sobel, Huitzu Tu  and Hai-Bo Yu
for discussions, and especially to Jonathan Feng, John Learned and
Louis Strigari.  This work was supported by NSF
grants PHY--0239817,0314712,0551164 and 0653656, DOE grant
DE-FG02-04ER41291, and the Alfred P.~Sloan Foundation.  JK is grateful
to CERN and the organizers of Strings '08, where part of this work was
done, and to the organizers of Dark 09 for their hospitality.

\bibliographystyle{ws-procs9x6}

\begin{thebibliography}{99}

\bibitem{Bernabei:2008yi}
  R.~Bernabei {\it et al.}, [DAMA Collaboration],
  arXiv:0804.2741 [astro-ph];
  Riv.\ Nuovo Cim.\  {\bf 26N1}, 1 (2003);
  Int.\ J.\ Mod.\ Phys.\  D {\bf 13}, 2127 (2004).

\bibitem{Drukier:1986tm}
  A.~K.~Drukier, K.~Freese and D.~N.~Spergel,
  Phys.\ Rev.\  D {\bf 33}, 3495 (1986);
  K.~Freese, J.~A.~Frieman and A.~Gould,
  Phys.\ Rev.\  D {\bf 37}, 3388 (1988).

\bibitem{Angloher:2002in}
  G.~Angloher {\it et al.},
  Astropart.\ Phys.\  {\bf 18}, 43 (2002).

\bibitem{Akerib:2005kh}
  D.~S.~Akerib {\it et al.}, [CDMS Collaboration],
  Phys.\ Rev.\ Lett.\  {\bf 96}, 011302 (2006);
  Z.~Ahmed {\it et al.}, [CDMS Collaboration],
  arXiv:0802.3530 [astro-ph].

\bibitem{Angle:2007uj}
  J.~Angle {\it et al.}, [XENON Collaboration],
  Phys.\ Rev.\ Lett.\  {\bf 100},021303(2008).

\bibitem{Lin:2007ka}
  S.~T.~Lin {\it et al.}, [TEXONO Collaboration],
  arXiv:0712.1645 [hep-ex].

\bibitem{Avignone:2008xc}
  F.~T.~Avignone, P.~S.~Barbeau and J.~I.~Collar,
  arXiv:0806.1341 [hep-ex].

\bibitem{Aalseth:2008rx}
  C.~E.~Aalseth {\it et al.},
  arXiv:0807.0879 [astro-ph].

\bibitem{Brhlik:1999tt}
See, \eg,
  M.~Brhlik and L.~Roszkowski,
  Phys.\ Lett.\  B {\bf 464}, 303 (1999);
  P.~Belli, R.~Bernabei, A.~Bottino, F.~Donato, N.~Fornengo,
   D.~Prosperi and S.~Scopel,
  Phys.\ Rev.\  D {\bf 61}, 023512 (2000).

\bibitem{Gondolo:2005hh}
  P.~Gondolo and G.~Gelmini,
  Phys.\ Rev.\  D {\bf 71}, 123520 (2005).

\bibitem{Bernabei:2007hw}
  R.~Bernabei {\it et al.},
  Eur.\ Phys.\ J.\  C {\bf 53}, 205 (2008).

\bibitem{Petriello:2008jj}
  F.~Petriello and K.~M.~Zurek,
  arXiv:0806.3989 [hep-ph].

\bibitem{Chang:2008xa}
  S.~Chang, A.~Pierce and N.~Weiner,
  arXiv:0808.0196 [hep-ph].

\bibitem{Fairbairn:2008gz}
  M.~Fairbairn and T.~Schwetz,
  arXiv:0808.0704 [hep-ph].

\bibitem{Savage:2008er}
  C.~Savage, G.~Gelmini, P.~Gondolo and K.~Freese,
  arXiv:0808.3607 [astro-ph].

\bibitem{Bottino:2003iu}
  A.~Bottino, F.~Donato, N.~Fornengo and S.~Scopel,
  Phys.\ Rev.\  D {\bf 68}, 043506 (2003);
  Phys.\ Rev.\  D {\bf 77}, 015002 (2008);
  arXiv:0806.4099 [hep-ph].

\bibitem{Smith:2001hy}
  D.~Tucker-Smith and N.~Weiner,
  Phys.\ Rev.\  D {\bf 64}, 043502 (2001);
  Phys.\ Rev.\  D {\bf 72}, 063509 (2005).

\bibitem{Feng:2008ya}
  J.~L.~Feng and J.~Kumar,
  Phys.\ Rev.\ Lett.\  {\bf 101}, 231301 (2008).

\bibitem{Feng:2008dz}
  J.~L.~Feng, J.~Kumar and L.~E.~Strigari,
  Phys.\ Lett.\  B {\bf 670}, 37 (2008).

\bibitem{Feng:2008mu}
  J.~L.~Feng, H.~Tu and H.~B.~Yu,
  JCAP {\bf 0810}, 043 (2008).

\bibitem{Foot:2008nw}
  R.~Foot,
  arXiv:0804.4518 [hep-ph].

\bibitem{Khlopov:2008ty}
  M.~Y.~Khlopov and C.~Kouvaris,
  arXiv:0806.1191 [astro-ph].

\bibitem{Andreas:2008xy}
  S.~Andreas, T.~Hambye and M.~H.~G.~Tytgat,
  arXiv:0808.0255 [hep-ph].

\bibitem{Dudas:2008eq}
  E.~Dudas, S.~Lavignac and J.~Parmentier,
  arXiv:0808.0562 [hep-ph].
  
\bibitem{Kim:2009ke}
  Y.~G.~Kim, K.~Y.~Lee and S.~Shin,
  JHEP {\bf 0805}, 100 (2008)
  [arXiv:0803.2932 [hep-ph]];
  Y.~G.~Kim and S.~Shin,
  arXiv:0901.2609 [hep-ph].

\bibitem{Feng:2008qn}
  J.~L.~Feng, J.~Kumar, J.~Learned and L.~E.~Strigari,
  arXiv:0808.4151 [hep-ph].

\bibitem{Gould:1987ir}
  A.~Gould,
  Astrophys.\ J.\  {\bf 321}, 571 (1987).

\bibitem{Gould:1987ju}
  A.~Gould,
  Astrophys.\ J.\  {\bf 321}, 560 (1987).

\bibitem{Hooper:2008cf}
  D.~Hooper, F.~Petriello, K.~M.~Zurek and M.~Kamionkowski,
  arXiv:0808.2464 [hep-ph].

\bibitem{Griest:1986yu}
  K.~Griest and D.~Seckel,
  Nucl.\ Phys.\  B {\bf 283}, 681 (1987)
  [Erratum-ibid.\  B {\bf 296}, 1034 (1988)].

\bibitem{Kamionkowski:1994dp}
  M.~Kamionkowski, K.~Griest, G.~Jungman and B.~Sadoulet,
  Phys.\ Rev.\ Lett.\  {\bf 74}, 5174 (1995).

\bibitem{Jungman:1994jr}
  G.~Jungman and M.~Kamionkowski,
  Phys.\ Rev.\  D {\bf 51}, 328 (1995).

\bibitem{Desai:2004pq}
  S.~Desai {\it et al.}  [Super-Kamiokande Collaboration],
  Phys.\ Rev.\  D {\bf 70}, 083523 (2004)
  [Erratum-ibid.\  D {\bf 70}, 109901 (2004)].

\bibitem{Montaruli:1999nv}
  T.~Montaruli  [MACRO Collaboration],
  arXiv:hep-ex/9905020.

\bibitem{delosHeros:2007hy}
  C.~de los Heros, et al.,
  arXiv:astro-ph/0701333.

\bibitem{Jungman:1995df}
  G.~Jungman, M.~Kamionkowski and K.~Griest,
  Phys.\ Rept.\  {\bf 267},195(1996).

\bibitem{Ashie:2005ik}
  Y.~Ashie {\it et al.}  [Super-Kamiokande Collaboration],
  Phys.\ Rev.\  D {\bf 71}, 112005 (2005).

\bibitem{Cheng:1988im}
  H.~Y.~Cheng,
  Phys.\ Lett.\  B {\bf 219}, 347 (1989).

\bibitem{Ellis:2001hv}
  J.~R.~Ellis, J.~L.~Feng, A.~Ferstl, K.~T.~Matchev and K.~A.~Olive,
  Eur.\ Phys.\ J.\  C {\bf 24}, 311 (2002).





\end{thebibliography}


\end{document}